\documentclass{ws-ijmpa}
\usepackage{overcite}

\begin{document}

\markboth{S. Ohkubo}
{Alpha clustering  and weak coupling in the A=90 region }

%
\catchline{}{}{}{}{}
%

\title{ALPHA CLUSTERING AND WEAK COUPLING IN THE  $^{93}$Nb REGION 
}

\author{S. Ohkubo\footnote{shigeo@cc.kochi-wu.ac.jp
}}

\address{Department of Applied Science, Kochi Women's University, Kochi 780-8515,
 Japan  and \\ Research Center for Nuclear Physics, Osaka University, Ibaraki, 
Osaka 567-0047, Japan 
}

\maketitle

\begin{history}
\received{\today }
\end{history}

\begin{abstract}
From the viewpoint of a unified description of  cluster structure and scattering in the 
 $A=90$ region,  $\alpha$ scattering from $^{89}$Y  is investigated.     $\alpha$ clustering  and 
weak coupling  in $^{93}$Nb  is discussed.
\end{abstract}


\ccode{PACS numbers: 21.60.Gx,24.10.Ht,25.55.Ci,27.60.+j}

\section{Introduction}
 It has been  widely accepted that the $\alpha$ cluster structure exists  in
 light nuclei.\cite{Hiura1972,Fujiwara1980}  It has been 
also shown that the $\alpha$ cluster 
picture persists even in the  $fp$-shell region like $^{44}$Ti where the 
spin-orbit force becomes strong.\cite{Michel1998,Yamaya1998,Sakuda1998}

 Clustering is realized  due to
 the logic that  the interaction of subunit clusters in nuclei is 
 {\it internally (intracluster)  strong but  externally (intercluster) 
 weak}.\cite{Hiura1972}  This is most typically seen in 
 $^{8}$Be with the $\alpha$+$\alpha$
 cluster structure. This logic of {\it internally strong but externally weak}
 is  justified by a microscopic model  study  which uses a two-body nuclear
 force and  fully takes into account the Pauli principle.\cite{Hiura1972}

  In the $p$-shell closure region, $^{16}$O$\sim$ $^{20}$Ne,  a weak coupling model was proposed phenomenologically 
 \cite{Arima1967} to account
 for the similarity of the structure of the rotational band built on the mysterious
0$^+$ state of $^{16}$O at 6.06 MeV  and the intruder  band   in $^{19}$F     starting 
from the $\frac{1}{2}^-$ state at 0.11 MeV to the ground band of 
$^{20}$Ne.   Nemoto and Band\={o}
 \cite{Nemoto1972} clarified in the microscopic model  that this weak coupling is 
a  consequence of both the development of  $\alpha$
clustering  and the Pauli principle, which   
 causes  internally strong 
and externally weak binding of the clusters.

 The persistency of this weak coupling
due to  $\alpha$ clustering in the $sd$-shell closure region,  $^{40}$Ca $\sim^{44}$Ti, was also
 shown.\cite{Sakuda1998}
 It is  important to know whether this   weak coupling  is universal and persists in much 
heavier nuclei where the 
closed shells for protons and neutrons are different. It is the purpose of this paper
 to study whether the weak coupling picture due to $\alpha$ clustering still
 persists in the $A=90$ region.

\section{ $\alpha$ clustering in the $A=90$  region }
 The structure of nuclei in the $A=90$ region has been mostly studied in 
the shell model and many properties have been successfully 
explained.\cite{Vervier1966,Ball1972,Gloeckner1975}   
However, for some nuclei the existence of intruder states, which are 
difficult in the shell model,  have been suggested. For example, in the typical nucleus
$^{94}$Mo, which has two protons and two neutrons outside the double 
closed shell, the shell model calculation needs a  large effective charge
 to explain the large $B(E2)$ values of the ground band.

 On the other hand, from the viewpoint of a unified description of bound and scattering for 
the $\alpha$+$^{90}$Zr system, the present author suggested  that the $\alpha$
 cluster model is still successful in understanding the structure of 
$^{94}$Mo.\cite{Ohkubo1995}   The structure of  $^{94}$Mo  is also  described  
in the
$\alpha$+$^{90}$Zr cluster model  using a phenomenological potential
 \cite{Buck1995,Atzrott1996,Michel2000,Souza2005} 
 and  a double folding model potential.\cite{Atzrott1996} 

\begin{figure}[pb]
\centerline{\psfig{file=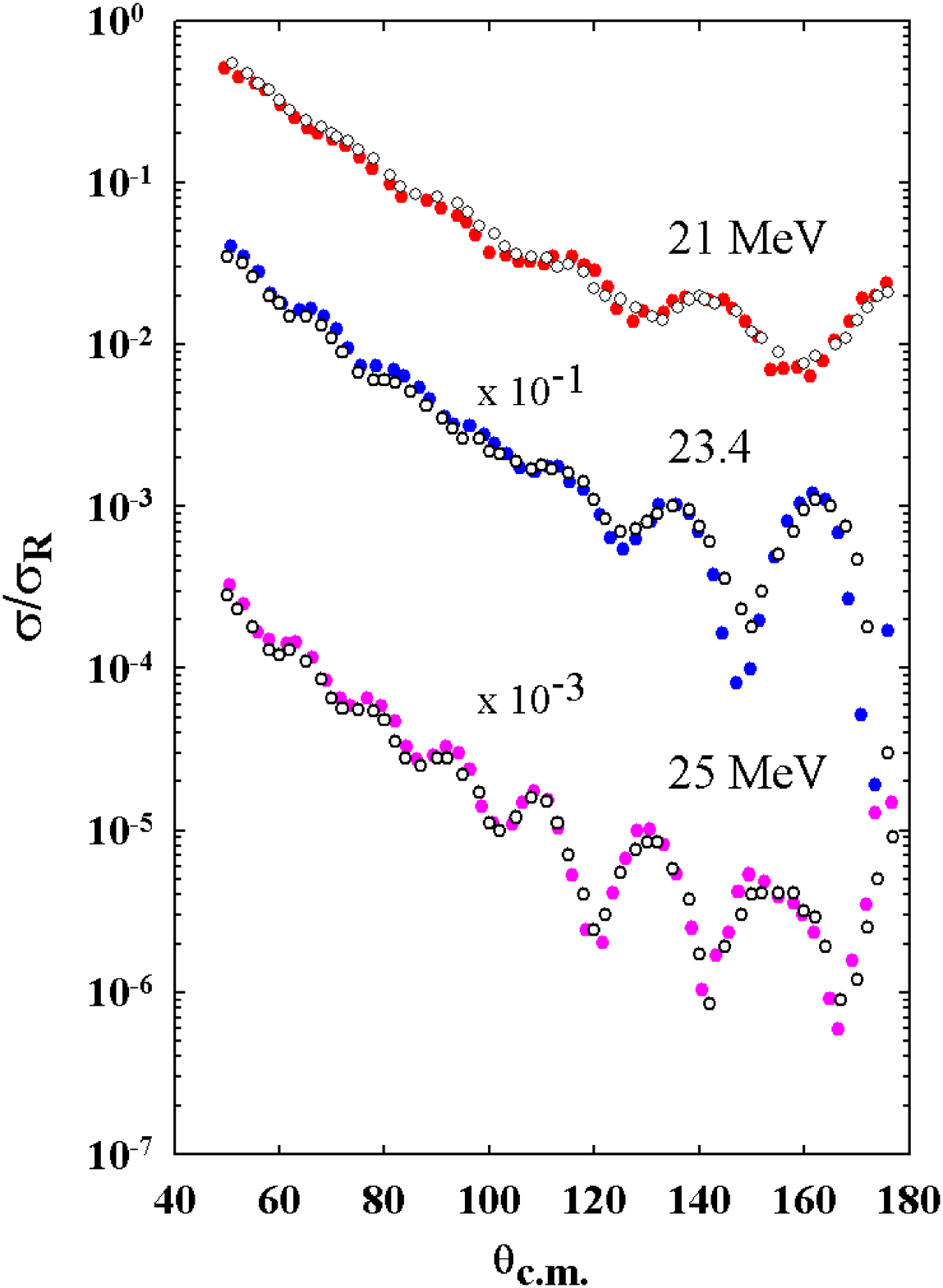,width=4.7cm}}
\vspace*{8pt}
\protect\caption{(Color online) The similarity  of the experimental angular distributions  of 
 $\alpha$+$^{89}$Y (filled circles) and  $\alpha$+$^{90}$Zr scattering (open circles) 
at $E_\alpha$=21, 23.4 and 25 MeV
 \cite{Wit1975} 
is shown.
 \label{fig1}}
\end{figure}


Although there are not many experimental data of $\alpha$ transfer reactions
in this region compared with the $^{20}$Ne and $^{44}$Ti regions, there are
some indications which suggest the importance of $\alpha$ clustering in this
 mass region.
 For example, Moln\'{a}r {\it et al.} \cite{Molnar1986} suggested that
 the intruder band  states 0$^+$(1.581 MeV), 2$^+$ (2.225 MeV) and 4$^+$
 (2.857 MeV) in $^{96}$Zr have four-particle four-hole character  just as in $^{16}$O. 
In the $^{90}$Zr$\sim$$^{94}$Mo region $\alpha$  transfer reactions, which 
 were very powerful in light nuclei in selectively populating  the $\alpha$ cluster 
states, were reported
 \cite{Fulbright1977,Fulbright1979,Umeda1984,Berg1983,Yamaya1998}. However, the 
$\alpha$ cluster states have not been clearly identified. 
Therefore another approach  from the viewpoint of $\alpha$ clustering 
seems   necessary.

\par

\begin{figure}[pb]
\centerline{\psfig{file=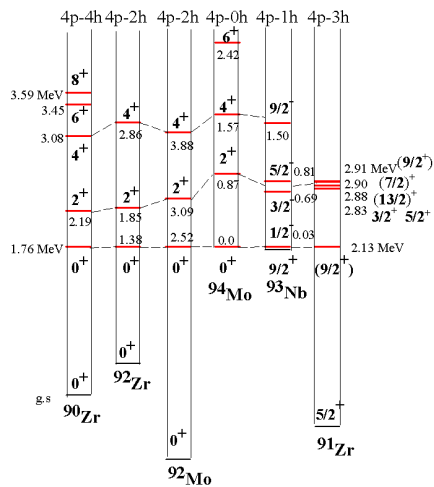,width=6.0cm}}
\vspace*{8pt}
\caption{(Color online) Experimental  energy levels \cite{Energy,Orce2007} of  the candidates for 
 the $\alpha$ cluster (4p-nh) states with the weak coupling feature
 in the $^{90}$Zr$\sim$$^{94}$Mo region are displayed connected  by the dashed lines 
so that the band head states are in coincidence. Excitation energies are given in MeV.
 \label{fig2}}
\end{figure}


 A unified description  of $\alpha$ cluster  bound states and 
$\alpha$ particle scattering  may  be one of the useful approaches in this region.
In fact, in the $^{44}$Ti region where $\alpha$ transfer reactions were not
sufficiently available, a unified description of bound states and $\alpha$
particle scattering from $^{40}$Ca using a global potential was  powerful
in revealing the  $\alpha$ clustering aspects  in  $^{44}$Ti and 
 $^{40}$Ca.\cite{Michel1998} 
 Because the structure of the typical nucleus $^{94}$Mo has been shown to be
described in the $\alpha$ cluster model, it seems useful to take this approach
for the nuclei in the $^{90}$Zr$\sim$$^{94}$Mo region.
In fact, as displayed in Fig.~1, the observed angular distributions in $\alpha$
 particle scattering from $^{89}$Y at $E_\alpha$=21-25 MeV are very similar to those for 
$\alpha$+$^{90}$Zr scattering. The observed excitation function
at $\theta_{\rm c.m.}$=176$^\circ$ for $\alpha$+$^{89}$Y scattering
 shows a    deep minimum at around $E_\alpha$=23.4 MeV similar to that for $\alpha$+$^{90}$Zr
 scattering.\cite{Wit1975}
 Furthermore, as seen in Fig.~2, the band structures built on the 4p-nh states resemble 
each other as was the case in the $^{16}$O$\sim$$^{20}$Ne region.\cite{Nemoto1972}

The above  two facts, the close angular distributions in  the scattering (Fig.~1) and the energy levels   with a 
similar band structure (Fig.~2),   suggest that 
 the potentials for the  $\alpha$+$^{89}$Y and the $\alpha$+$^{90}$Zr 
 systems are  very similar.
As the  $\alpha$+$^{90}$Zr structure in  
$^{94}$Mo is an analog of  the $\alpha$+$^{16}$O  cluster structure in $^{20}$Ne
 and the $\alpha$+$^{40}$Ca  cluster structure in $^{44}$Ti,
 the  $\alpha$+$^{89}$Y structure in $^{93}$Nb may be regarded  an analog of the
  $\alpha$+$^{15}$N  structure in $^{19}$F in the $sd$-shell and the
  $\alpha$+$^{39}$K  structure in $^{43}$Sc in the $fp$-shell.
Therefore, to clarify whether the weak coupling feature persists in the  heavier  
$A=90$ region
 it seems  useful to study the   $\alpha$+$^{89}$Y cluster  structure of $^{93}$Nb 
in comparison with the   $\alpha$+$^{90}$Zr structure of $^{94}$Mo.
Recently Kiss {\it et al.} \cite{Kiss2008} extended the measurement 
of  $\alpha$+$^{89}$Y scattering  to the  lower energies, $E_\alpha$=16.2 and 19.4 MeV.
Also a  large $B(E2)$ value of the electric transition in $^{93}$Nb has been
observed  by Orce {\it et al.}\cite{Orce2007}

\begin{figure}[pb]
\centerline{\psfig{file=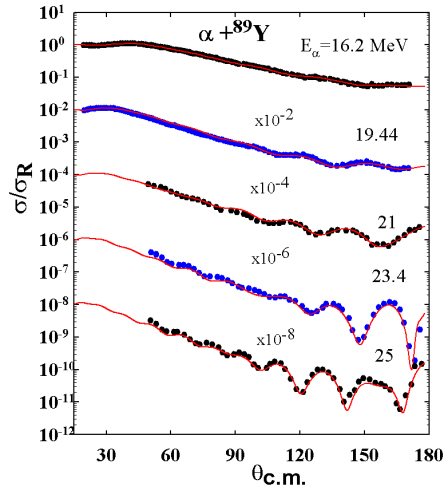,width=6cm}}
\vspace*{8pt}
\caption{(Color online) Calculated angular distributions in  
$\alpha$+$^{89}$Y
scattering  (solid lines) are compared with the experimental data (circles)
 \cite{Wit1975,Kiss2008}.
 \label{fig3}}
\end{figure}

\section{Analysis of $\alpha$ particle scattering from $^{89}$Y }


  We analyze the $\alpha$+$^{89}$Y scattering by taking a Woods-Saxon squared phenomenological 
potential,
  which is similar to the one  used for the $\alpha$+$^{90}$Zr system.\cite{Ohkubo1995,Michel2000}
The form factor of the optical potential is given in the standard notation as follow\cite{Michel2000}:

\begin{equation}
V(r)=-V_0\{1+\alpha \exp[-(r/\rho)^2]\}/
\{1+\exp[(r-R_V)/2a_V]\}^2+V_C(r), \\
\end{equation}
\begin{equation}
W(r)=-W_0/\{1+\exp[(r-R_W)/2a_W]\}^2, 
\end{equation}
for the real and imaginary potentials, respectively.
The Coulomb potential $V_C(r)$ is assumed to be a uniformly charged sphere with a radius 
$R_c$=7.49 fm. 
The real potential  parameters $V_0$=35 MeV, $R_V$=7.49 fm, $a_V=$0.43 fm, $\rho$=4.86 fm  and 
$\alpha$=4.748
 are fixed for all the incident energies.  The $V_0$,  $a_V$ and $\alpha$ parameters 
are the same as
those for  the  $\alpha$+$^{90}$Zr system, and the $R_V$ and $\rho$ parameters are  
scaled from those for the
 $\alpha$+$^{90}$Zr system.\cite{Michel2000}
The imaginary 
potential  parameters were adjusted to fit the data and are given in Table I.

As shown in Fig.~3, a good agreement is  obtained with the experimental data. 
The characteristic deep minimum in the angular distribution at the extreme backward angle
at $E_\alpha$=23.4 MeV, which is also seen in the $\alpha$+$^{90}$Zr scattering (Fig.~1), is well reproduced by the 
calculation. As discussed in the $\alpha$+$^{90}$Zr scattering \cite{Michel2000}, the oscillatory structure
 of the angular distribution at the backward angular region
 and the deep minimum at the extreme backward angle $\theta_{\rm c.m.}$=176$^\circ$ at $E_\alpha$=23.4 MeV 
are caused by the interference between the internal waves, which penetrate deep into the internal
 region of the potential and the barrier waves, which are reflected at  the surface region of the potential.
This shows that absorption is  incomplete and transparent for the  $\alpha$+$^{89}$Y scattering in this energy
 region and the potential  can be checked not only at the surface region but also in the internal region.
This means that the real part of the obtained potential may be used for the $\alpha$+$^{89}$Y cluster 
structure calculation in $^{93}$Nb.

\begin{table}[ph]
\tbl{Imaginary potential parameters used in the analysis of the angular distributions 
in $\alpha$ + $^{89}$Y
 scattering. }
{\begin{tabular}{lccl} \toprule 
$E_{\alpha}$ (MeV)&   $ W_0$ (MeV) &  $ R_W $  (fm)  & $a_W$ (fm) \\
 \colrule
16.2 & 14.5  & 7.04   & 0.162 \\
 19.44& 11.6  & 7.71  &  0.135 \\
 21   & 16.1  & 6.47  &  0.08 \\
 23.4 & 17.0  & 6.44  &  0.135 \\
  25  & 17.6  & 6.44  &  0.135 \\
\botrule
\end{tabular} \label{ta1}}
\end{table}

\section{$\alpha$ clustering in $^{93}$Nb and weak coupling  }

\begin{figure}[pb]
\centerline{\psfig{file=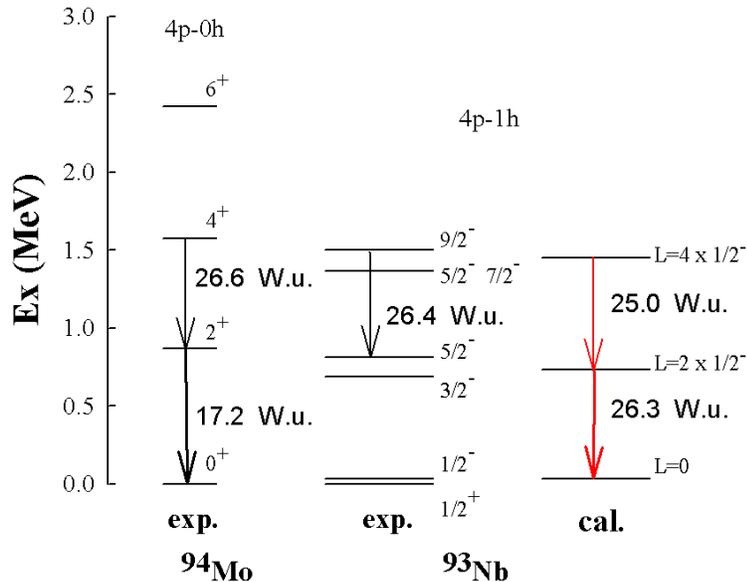,width=10cm}}
\vspace*{8pt}
\caption{(Color online) Calculated energy levels and $B(E2)$ values of $^{93}$Nb are compared with
the experimental data \cite{Energy,Orce2007}.
 The experimental $B(E2)$ values  for $^{94}$Mo \cite{Adamides1981} are shown for comparison. 
 \label{fig4}}
\end{figure}

As shown in Fig.~4, in $^{93}$Nb the first excited state $\frac{1}{2}^-$ appears at an extremely 
low excitation energy, $E_x$=0.031 MeV
above the ground state $\frac{9}{2}^+$  similar to the $\frac{1}{2}^-$  state at 0.11 MeV 
with  a well-developed $\alpha$+$^{15}$N cluster structure in $^{19}$F.
 In the $\alpha$ cluster model  this state may be understood  qualitatively 
that  an $\alpha$ particle is orbiting around   the core nucleus $^{89}$Y($\frac{1}{2}^-$) with an
 orbital angular momentum 
$L=0$. The $\frac{3}{2}^-$ state at $E_x$=0.687 MeV and the  $\frac{5}{2}^-$ state at $E_x$=0.810 MeV are considered
to be  doublet states, in which the $\alpha$ particle is orbiting around the core with $L=2$.  Recently 
Orce {\it et al.} \cite{Orce2007} observed a strong $E2$ transition from the  $\frac{9}{2}^-$ state
 at $E_x$=1.500 MeV 
to the $\frac{5}{2}^-$ state ($E_x$=0.810 MeV) with $B(E2$)=26.4$^{+9.7}_{-6.2}$ W.u. 
This state  may be considered to be a partner of the doublet states, in which  the $\alpha$ particle is 
 coupled to  the core with $L=4$. The state at 1.364 MeV is reported to be $\frac{5}{2}^-$ or
 $\frac{7}{2}^-$.\cite{Energy}

The calculated energy levels and the $B(E2)$ values are displayed in Fig.~4 in 
comparison with the experimental data.\cite{Energy,Orce2007}  
In the  calculation of  the $\alpha$ cluster structure we take  the real potential parameters used 
in the analysis of $\alpha$ + $^{89}$Y  scattering. To take  account of the
angular momentum dependence of the real potential, as discussed in Ref.[\citen{Ohkubo1995}],  we assume 
$\alpha$=$\alpha_0$-$c$$L$.  $\alpha_0$=4.17 is  fine tuned to reproduce the experimental threshold energy 
and   $c$=0.005 is  used, which is very small, consistent with the values needed in the $^{44}$Ti and 
$^{94}$Mo cases.\cite{Ohkubo1995} No spin-orbit potential is introduced.  The calculated  $N=16$ band states,  
which satisfy the Wildermuth condition $N$=$2n+L$=16 with $n$  being the  number of nodes 
 of the relative wave function,  fall  well near the experimental
 states.   The agreement of the calculated
 energy levels with the experimental data is good. The experimental $B(E2)$ value of the $E2$ transition
 from the  $\frac{9}{2}^-$ state (1.500 MeV) to the $\frac{5}{2}^-$ state (0.810 MeV)  is reproduced well
with a small  effective charge $\delta$e=0.2, which is consistent with the value 
in Ref.[\citen{Ohkubo1995}].  
In Fig.~4, we note that the calculated and observed  $B(E2$) values for $^{93}$Nb are comparable to the observed values
 for $^{94}$Mo. 
Thus it is found that the $\alpha$ cluster model works  in $^{93}$Nb  in addition to $^{94}$Mo
and the structure of the energy levels and the $B(E2)$ values suggests that weak coupling picture due
to $\alpha$ clustering persists in this mass region.
 The  $\alpha$-hole energy estimated from the experimental binding energies of $^{94}$Mo, $^{90}$Zr,
$^{93}$Nb,  $^{89}$Y and the $\alpha$ particles is 0.17 MeV 
  for a proton hole in the $2p_\frac{1}{2}$ shell. This is still small compared with 0.86 MeV
of the $\alpha$-hole energy in $^{19}$F.\cite{Arima1967}
It is desired to observe the $E2$ transition from the
 $\frac{5}{2}^-$ state (0.810 MeV) to the $\frac{1}{2}^-$ (0.031 MeV) and to compare with the 
predicted value. 

\section{Summary}
 We have studied $\alpha$+$^{89}$Y scattering and the $\alpha$+$^{89}$Y cluster structure in $^{93}$Nb
in a unified way by using a phenomenological potential. It is found that the potential which reproduces the 
angular distributions in $\alpha$+$^{89}$Y scattering is very similar to that for the $\alpha$+$^{90}$Zr system. The calculated
 energy levels and $B(E2)$ values obtained using the  potential reproduce  the 
experimental data well.
 The weak coupling picture due to $\alpha$ clustering
seems to persist in $^{93}$Nb.  A systematic measurement  of the $B(E2)$ values of the  nuclei 
in the  $^{90}$Zr $\sim$ $^{94}$Mo region  is highly desired to study the persistency 
of the weal coupling feature due to 
$\alpha$ clustering.

\section*{Acknowledgments}
 The author    has been supported by a
 Grant-in-aid for Scientific Research  of the Japan Society for Promotion of Science (No. 16540265)
 and the Yukawa Institute for Theoretical Physics (GCOE) where this work has been completed.

\end{document}